\documentclass{amsart}
\usepackage{latexsym}
\usepackage{amsfonts}
\usepackage{amssymb}
\usepackage{graphicx}
\usepackage{graphicx}

\newtheorem{theorem}{Theorem}[section]
\newtheorem{corollary}[theorem]{Corollary}

\newtheorem{proposition}[theorem]{Proposition}
\theoremstyle{definition}
\newtheorem{definition}{Definition}[section]

\theoremstyle{remark}
\newtheorem{remark}{Remark}[section]
\numberwithin{equation}{section}

\newcommand{\Lg}{\mathfrak{g}}
\newcommand{\Lh}{\mathfrak{h}}
\newcommand{\Lk}{\mathfrak{k}}
\newcommand{\Lp}{\mathfrak{p}}
\newcommand{\La}{\mathfrak{a}}
\newcommand{\Ld}{\mathfrak{d}}
\newcommand{\Lb}{\mathfrak{b}}
\newcommand{\Ll}{\mathfrak{l}}
\newcommand{\Lm}{\mathfrak{m}}
\newcommand{\Ln}{\mathfrak{n}}
\newcommand{\Lt}{\mathfrak{t}}
\newcommand{\Lu}{\mathfrak{u}}
\newcommand{\Lo}{\mathfrak{o}}
\newcommand{\Ly}{\mathfrak{y}}
\newcommand{\gl}{\mathfrak{gl}}
\newcommand{\Lsl}{\mathfrak{sl}}
\newcommand{\Lsp}{\mathfrak{sp}}
\newcommand{\Lsu}{\mathfrak{su}}
\newcommand{\Lso}{\mathfrak{so}}

\newcommand{\ad}{\mathrm{ad}}
\newcommand{\Ad}{\mathrm{Ad}}
\newcommand{\RE}{\mathrm{Re}}
\newcommand{\IM}{\mathrm{Im}}
\newcommand{\A}{\mathrm{A}}
\newcommand{\z}{\mathrm{z}}
\newcommand{\diag}{\mathrm{diag}}
\newcommand{\tr}{\mathrm{tr}}
\newcommand{\ii}{\mathbf{i}}
\newcommand{\jj}{\mathbf{j}}
\newcommand{\kk}{\mathbf{k}}
\newcommand{\ee}{\mathbf{e}}
\newcommand{\KK}{\mathbb{K}}
\newcommand{\ZZ}{\mathbb{Z}}
\newcommand{\NN}{\mathbb{N}}
\newcommand{\RR}{\mathbb{R}}
\newcommand{\CC}{\mathbb{C}}
\newcommand{\HH}{\mathbb{H}}
\newcommand{\HC}{\HH_{\CC}}
\newcommand{\D}{\displaystyle}

\begin{document}

\title[]
{A Generalization of random matrix ensemble I: general theory$^1$}
\author[]{Jinpeng An}
\address{School of mathematical science, Peking University, Beijing, 100871, P. R. China}
\email{anjinpeng@math.pku.edu.cn}

\author[]{Zhengdong Wang}
\address{School of mathematical science, Peking University, Beijing, 100871, P. R. China }
\email{zdwang@pku.edu.cn}

\author[]{Kuihua Yan}
\address{School of mathematics and physics, Zhejiang Normal
University, Zhejiang Jinhua, 321004, P. R. China }
\email{yankh@zjnu.cn}

\begin{abstract}
We give a generalization of the random matrix ensembles, including
all classical ensembles. Then we derive the joint density function
of the generalized ensemble by one simple formula, which give a
direct and unified way to compute the density functions for all
classical ensembles and various kinds of new ensembles. An
integration formula associated with the generalized ensemble is
also given. We also give a classification scheme of the
generalized ensembles, which will include all classical ensembles
and some new ensembles which were not considered before.
\end{abstract}
\maketitle


\footnotetext[1]{This work is supported by the 973 Project
Foundation of China ($\sharp$TG1999075102).
}

\vskip 0.5cm
\section{Introduction}

\vskip 0.5cm One of the most fundamental problems in the theory of
random matrices is to derive the joint density functions for the
eigenvalues (or equivalently, the measures associated with the
eigenvalue distributions) of various types of matrix ensembles. In
his monograph \cite{Me}, Mehta summarized the classical analysis
methods by which the density functions for various types of
ensembles were derived case by case. But a systematical method to
compute the density functions was desired.

\vskip 0.3cm The first achievement in this direction was made by
Dyson \cite{Dy}, who introduced an idea of expressing various
kinds of circular ensemble in terms of symmetric spaces with
invariant probability measures. From then on, guided by Dyson's
idea, many authors observed new random matrix ensembles in terms
of Cartan's classification of Riemannian symmetric spaces, and
obtained the joint density functions for such ensembles using the
integration formula on symmetric space (see, for example,
\cite{AZ,Ca1,Ca2,Du,Iv,TBFM,Zi}). Here we mention the recent work
of Due\~{n}ez \cite{Du} briefly. Due\~{n}ez explored the random
matrix ensembles which correspond to infinite families of compact
irreducible Riemannian symmetric spaces of type I, including
circular orthogonal and symplectic ensembles and various kinds of
Jacobi ensembles. Using an integration formula associated with the
$KAK$ decomposition of compact groups, he obtained the induced
measure on the space of eigenvalues associated with the underlying
symmetric space, and then derived the eigenvalue distribution of
the corresponding random matrix ensemble. These methods of
deriving the eigenvalue distributions of random matrix ensembles
by means of Riemannian symmetric spaces were summarized by the
excellent review article of Caselle and Magnea \cite{CM}.

\vskip 0.3cm In this paper we provide a generalization of the
random matrix ensembles, including all classical ensembles, and
then give an unified way to derive the joint density function for
the eigenvalue distribution by one simple formula. The proof of
this formula make no use of integration formula. In fact, the
corresponding integration formula can be derived from this formula
as corollary. We also give a classification scheme of the
generalized random matrix ensembles, which will include all
classical ensembles and some new ensembles which were not
considered before.

\vskip 0.3cm More precisely, Let $\sigma:G\times X\rightarrow X$
be a smooth action of a Lie group $G$ on a Riemannian manifold
$X$, preserving the induced Riemannian measure $dx$. Let $p(x)$ be
a $G$-invariant smooth function on $X$, and consider the measure
$p(x)dx$ on $X$, which is not necessary a finite measure. We
choose a closed submanifold $Y$ of $X$ consisting of
representation points for almost all $G$-orbits in $X$. The
Riemannian structure on $X$ induces a Riemannian measure $dy$ on
$Y$. Let $K$ be the closed subgroup of $G$ which fixes all points
in $Y$, then the map $\sigma$ reduces to a map $\varphi:G/K\times
Y\rightarrow X$. Suppose there is a $G$-invariant measure $d\mu$
on $G/K$, and suppose $\dim(G/K\times Y)=\dim X$, then it can be
proved that the pull back measure $\varphi^*(p(x)dx)$ of the
measure $p(x)dx$ is of the form $\varphi^*(p(x)dx)=d\mu d\nu$ for
some measure $d\nu$ on $Y$, which is just the measure associated
with the eigenvalue distribution. The measure $d\nu$ can be
expressed as the form $d\nu(y)=\mathcal{P}(y)dy$ for some function
$\mathcal{P}(y)$ on $Y$, which is just the joint density function.
We write $\mathcal{P}(y)$ as the form $\mathcal{P}(y)=p(y)J(y)$,
then under some orthogonality condition (that is $T_yY\perp
T_yO_y$ for almost all $y\in Y$), we can compute the factor $J(y)$
by the following formula
\begin{equation}\label{E:main}
J(y)=C|\det\Psi_y|,
\end{equation}
where $C$ is a constant, which can also be computed explicitly.
This formula is the main result of this paper, the density
function $\mathcal{P}(y)$ and the eigenvalue distribution $d\nu$
are determined by it. Here the map $\Psi_y:\Ll\rightarrow T_yO_y$
is defined by
$$
\Psi_y(\xi)=\frac{d}{dt}\Big|_{t=0}\hskip 0.1cm \sigma_{\exp
t\xi}(y),
$$
where $\Ll$ is a linear subspace of the Lie algebra $\Lg$ of $G$
such that $\Lg=\Lk\oplus\Ll$, $\Lk$ is the Lie algebra of $K$. We
call the system $(G,\sigma,X,p(x)dx,Y,dy)$ a \emph{generalized
random matrix ensemble}. The measure $d\nu$ and the function
$\mathcal{P}(y)$ on $Y$ are called \emph{generalized eigenvalue
distribution} and \emph{generalized joint density function},
respectively. Using Formula \eqref{E:main}, one can derive the
joint density function for Gaussian ensemble, chiral ensemble, new
transfer matrix ensembles, circular ensemble, Jacobi ensemble, and
some other new generalized ensembles. The precise deriving process
will be the content of a sequel paper \cite{AWY}. Here we should
point out that the proof of Formula \eqref{E:main} is not
difficult, but this formula is very effective and available. The
derivation of all concrete examples in \cite{AWY}, including all
classical random matrix ensembles, will be based on it.

\vskip 0.3cm Once the eigenvalue distribution $d\nu$ is derived by
Formula \eqref{E:main}, under a covering condition, we can get the
associated integration formula. The Weyl integration formula for
compact Lie groups, the Harish-Chandra's integration formula for
complex semisimple Lie groups and real reductive groups, the
integration formulae on Riemannian symmetric spaces of noncompact
and compact types which were appeared in \cite{He}, as well as
their Lie algebra versions are all of particular cases of it (see
\cite{AWY}).

\vskip 0.3cm Now let us give a sketch of each section of this
paper. In \S 2 we will develop some geometrical preliminaries on
the geometry of $G$-space which will be required to establish the
generalized ensemble. After presenting four conditions, that is,
the invariance condition, the transversality condition, the
dimension condition, and the orthogonality condition, on which the
definition of generalized ensemble will be based, we will prove in
Theorem \ref{T:det} a primary form of Formula \eqref{E:main}.

\vskip 0.3cm \S 3 will be devoted to the integration over
$G$-spaces, which will be needed when we derive the integration
formula associated with the generalized random matrix ensemble.
Based on the four conditions presented in \S 2 and a covering
condition, we will prove an integration formula in Theorem
\ref{T:Weylbianhuan}, which converts the integration over a
$G$-space to the integration by first integrating over each
$G$-orbit, and then integrating over the orbits space. Two
criterions on when the covering condition holds will also be
given.

\vskip 0.3cm Prepared by the preliminaries of \S 2 and \S 3, In \S
4 we will give the precise definition of the generalized random
matrix ensemble, as well as the associated generalized eigenvalue
distribution and generalized joint density function. In Theorem
\ref{T:distribution} the Formula \eqref{E:main} will be presented,
from which the associated eigenvalue distribution measure and
density function will be derived for various concrete examples of
the generalized ensemble in an unified way in \cite{AWY}.

\vskip 0.3cm In \S 5 we will give a classification scheme of
generalized ensembles, that is, the linear ensemble, the nonlinear
noncompact ensemble, the compact ensemble, the group and algebra
ensembles, as well as the pseudo-group and pseudo-algebra
ensembles. According to this classification scheme, Gaussian
ensemble and chiral ensemble are included in linear ensemble, new
transfer matrix ensembles is included in nonlinear noncompact
ensemble, circular and Jacobi ensembles are included in compact
ensemble. Some new ensembles which were not considered before will
also be included.

\vskip 0.5cm
\section{Geometry of $G$-spaces}

\vskip 0.5cm In this section we develop some geometrical
preliminaries which will be needed to establish our theory of the
generalized random matrix ensembles.

\vskip 0.3cm First we make some preparation about measures on
manifolds. Let $M$ be an $n$-dimensional smooth manifold. A
measure $dx$ on $M$ is called \emph{smooth} (or
\emph{quasi-smooth}) if on any local coordinate chart $(U;
x_1,\cdots,x_n)$ of $M$, $dx$ has the form $dx=f(x)dx_1\cdots
dx_n$, where $f$ is a smooth function on $U$ and $f>0$ (or $f\geq
0$), $dx_1\cdots dx_n$ is the Lebesgue measure on $\mathbb{R}^n$.
Note that the smooth measures on $M$ are unique up to multiplying
a positive smooth function on $M$, so the concept of set of
measure zero makes sense, which is independent of the choice of
smooth measure.

\vskip 0.3cm Let $M, N$ be two $n$-dimensional smooth manifolds,
and let $\varphi:M\rightarrow N$ be a smooth map. If $dy$ is a
smooth (or quasi-smooth) measure on $N$ which can be expressed
locally as $dy=f(y)dy_1\cdots dy_n$, we can define the \emph{pull
bake} $\varphi^*(dy)$ of $dy$ locally as
\begin{equation}\label{E:pullback}
\varphi^* (dy)=f(\varphi(x))\left|\det\left(\frac{\partial
y}{\partial x}\right)\right|dx_1\cdots dx_n.
\end{equation}
It is easily to check that the definition is compatible when we
choose different coordinate charts, and $\varphi^*(dy)$ is a
quasi-smooth measure on $M$. We can not expect $\varphi^*(dy)$ is
smooth in general, even if $dy$ is smooth, since $\varphi$ may
have critical points. But if $\varphi$ is a local diffeomorphism
and $dy$ is smooth, then $\varphi^*(dy)$ is smooth.

\vskip 0.3cm If $M, N$ are Riemannian manifolds and $dx, dy$ are
the associated Riemannian measures, then we can express the pull
back measure $\varphi^*(dy)$ globally. To do this, first we need
some comments on the ``determinant" of a linear map between two
different inner product vector spaces of the same dimension.
Suppose $V,W$ are two $n$-dimensional vector spaces with inner
products. For $n$ vectors $v_1,\cdots,v_n\in V$, let
$a_{ij}=\langle v_i,v_j\rangle$ for $1\leq i,j\leq n$, and define
$Vol(v_1,\cdots,v_n)=\sqrt{\det(a_{ij})}$. Note that if
$v_1,\cdots,v_n$ is an orthogonal basis, then
$Vol(v_1,\cdots,v_n)=|v_1|\cdots|v_n|$. For vectors in $W$ we
define the same things. Suppose $A:V\rightarrow W$ is a linear
map, define
\begin{equation}\label{E:det}
|\det A|=\frac{Vol(Av_1,\cdots,Av_n)}{Vol(v_1,\cdots,v_n)},
\end{equation}
where $v_1,\cdots,v_n$ is a basis of $V$. It is easily to check
that the definition is independent of the choice of the basis
$v_1,\cdots,v_n$. In the special case that $v_1,\cdots,v_n$ is an
orthogonal basis of $V$ and $Av_1,\cdots,Av_n$ are mutually
orthogonal, then
\begin{equation}\label{E:det2}
|\det A|=\frac{|Av_1|\cdots|Av_n|}{|v_1|\cdots|v_n|}.
\end{equation}
 Note that we can
only expect the norm of the determinant $|\det A|$ is well
defined, since the sign ``$\pm$" depends on the choice of
orientations of $V$ and $W$.

\begin{proposition}\label{P:pullback}
Suppose $M, N$ are two $n$-dimensional Riemannian manifolds with
the associated Riemannian measures $dx, dy$, repsectively. If
$\varphi:M\rightarrow N$ is a smooth map, then
\begin{equation}\label{E:Riemmannpullback}
\varphi^* (dy)=|\det(d\varphi)_x|dx.
\end{equation}
\end{proposition}

\begin{proof}
Suppose that in local coordinate charts the Riemannian metrics on
$M$ and $N$ are $ds^2=\Sigma_{ij}g_{ij}(x)dx_idx_j$ and
$d\widetilde{s}^2=\Sigma_{ij}\widetilde{g}_{ij}(y)dy_idy_j$,
respectively, where $g_{ij}(x)=\langle\frac{\partial}{\partial
x_i},\frac{\partial}{\partial x_j}\rangle$ and
$\widetilde{g}_{ij}(y)=\langle\frac{\partial}{\partial
y_i},\frac{\partial}{\partial y_j}\rangle$. Then by definition,
the Riemannian measures $dx,dy$ are
$dx=\sqrt{\det\big(g_{ij}(x)\big)}dx_1\cdots dx_n$ and
$dy=\sqrt{\det\big(\widetilde{g}_{ij}(y)\big)}dy_1\cdots dy_n$. We
have
\begin{align*}
|\det(d\varphi)_x|^2
=&\frac{Vol\left((d\varphi)_x(\frac{\partial}{\partial
x_1}),\cdots,(d\varphi)_x(\frac{\partial}{\partial
x_n})\right)^2}{Vol\left(\frac{\partial}{\partial
x_1},\cdots,\frac{\partial}{\partial x_n}\right)^2}\\
=&\frac{\det\left(\left\langle\sum_k\frac{\partial y_k}{\partial
x_i}(\frac{\partial}{\partial
y_k})_{\varphi(x)},\sum_l\frac{\partial y_l}{\partial
x_j}(\frac{\partial}{\partial
y_l})_{\varphi(x)}\right\rangle\right)}{\det\left(\left\langle\frac{\partial}{\partial
x_i},\frac{\partial}{\partial x_j}\right\rangle\right)}\\
=&\frac{\det\left(\sum_{kl}\frac{\partial y_k}{\partial
x_i}\frac{\partial y_l}{\partial
x_j}\widetilde{g}_{kl}(\varphi(x))\right)}{\det\big(g_{ij}(x)\big)}\\
=&\frac{\det\left(\left(\frac{\partial y_k}{\partial x_i}\right)^t
\Big(\widetilde{g}_{kl}(\varphi(x))\Big)\left(\frac{\partial
y_l}{\partial
x_j}\right)\right)}{\det\big(g_{ij}(x)\big)}\\
=&\frac{\left(\det\left(\frac{\partial y}{\partial
x}\right)\right)^2\det\big(\widetilde{g}_{ij}(\varphi(x))\big)}{\det\big(g_{ij}(x)\big)}.
\end{align*}
Hence
\begin{align*}
\varphi^*(dy)=&\sqrt{\det\big(\widetilde{g}_{ij}(\varphi(x))\big)}\left|\det\left(\frac{\partial
y}{\partial x}\right)\right|dx_1\cdots dx_n\\
=&|\det(d\varphi)_x|\sqrt{\det\big(g_{ij}(x)\big)}dx_1\cdots dx_n\\
=&|\det(d\varphi)_x|dx.
\end{align*}
\end{proof}

Now we come to the main geometric problems which will be concerned
in the following sections. Let $G$ be a Lie group, which acts on
an $n$-dimensional smooth manifold $X$. The action is denoted by
$\sigma: G\times X\rightarrow X$, and we denote
$\sigma_g(x)=\sigma(g,x)$. Our first goal is, roughly speaking, to
choose a representation point in each $G$-orbit $O_x = \{
\sigma_g(x): g\in G \}$, and the representation points should
depend smoothly on the orbits. But in general this aim can only be
achieved partially. So suppose we have a closed submanifold $Y$ of
$X$, which consists of the representation points of the orbits in
one's mind, such that $Y$ intersects ``almost all" orbits
transversally. More precisely, we suppose there are closed zero
measure subsets $X_\z \subset X$, $Y_\z \subset Y$. Let $X' = X
\setminus X_\z$, $Y' = Y \setminus Y_\z$, and suppose that

\vskip 0.3cm {\flushleft{\bf(a)}} \quad (\emph{invariance
condition}) \quad $X' = \D\bigcup_{y\in Y'} O_y$.

\vskip 0.3cm {\flushleft{\bf(b)}} \quad (\emph{transversality
condition}) \quad $T_y X=T_y O_y\oplus T_y Y$, \quad $\forall y\in
Y'$.

\vskip 0.3cm {\flushleft It} is clear that (a) implies $Y'=Y\cap
X'$, and then $Y_\z=Y\cap X_\z$. Notice that $X'$ and $Y'$ are
open and dense submanifolds of $X$ and $Y$, respectively. So
$\forall y\in Y'$, $T_y X' = T_y X$, $T_y Y' = T_y Y$.

\vskip 0.3cm Let $K=\{g\in G:\sigma_g(y)=y, \forall y\in Y\}$,
then $K$ is a closed subgroup of $G$. For $g\in G$, we denote
$[g]=gK$ in $G/K$. The map $\sigma: G\times X\rightarrow X$
reduces to a map $\varphi: G/K\times Y\rightarrow X$ by
$\varphi([g],y)=\sigma_g(y)$, and then induces a map $G/K\times
Y'\rightarrow X'$ by restriction, which we also denote by
$\varphi$. By the assumption above, $\varphi:G/K\times
Y'\rightarrow X'$ is surjective. For $x\in X$, Let $G_x=\{g\in
G:\sigma_g(x)=x\}$ be the isotropic subgroup associated with $x$.
Then $K\subset G_y, \forall y\in Y$. Let $dx, dy$ be smooth
measures on $X$ and $Y$, respectively. We suppose $dx$ is
$G$-invariant. In the following we suppose that

\vskip 0.3cm {\flushleft{\bf(c)}} \quad (\emph{dimension
condition}) \quad $\mathrm{dim}G_y=\mathrm{dim}K, \quad \forall
y\in Y'$.

\vskip 0.3cm {\flushleft This} means that $\forall y\in Y'$, $G_y$
and $K$ have the same Lie algebras, and the only difference
between $G_y$ and $K$ is that $G_y$ may have more components than
$K$. Then for some $y\in Y'$, we have
\begin{align*}
\mathrm{dim}X
=&\mathrm{dim}T_y X\\
=&\mathrm{dim}T_y Y+\mathrm{dim}T_y
O_y\\
=&\mathrm{dim}Y+\mathrm{dim}G-\mathrm{dim}G_y\\
=&\mathrm{dim}Y+ \mathrm{dim}G-\mathrm{dim}K .
\end{align*}
So $\varphi$ is a map between manifolds of the same dimension, and
the pull back $\varphi^*(dx)$ of $dx$ makes sense. Suppose also
that there is a $G$-invariant smooth measure $d\mu$ on $G/K$, then
the product measure $d\mu dy$ on $G/K\times Y$ is smooth, so
\begin{equation}\label{F:J(g,y)}
\varphi^*(dx)=J([g],y)d\mu dy
\end{equation}
for some $J\in C^\infty(G/K\times Y)$ with $J\geq 0$.

\begin{remark}
The $G$-invariant smooth measure $d\mu$ on $G/K$ exists if and
only if $\Delta_G|_K=\Delta_K$, where $\Delta_G$ and $\Delta_K$
are the modular functions on $G$ and $K$, respectively, see, e.g.,
Knapp \cite{Kn}, Section 8.3. For concrete examples in the
following sections, this condition always hold.
\end{remark}

\begin{proposition}\label{P:J}
The smooth function $J\in C^\infty(G/K\times Y)$ is independent of
the first variable $[g]\in G/K$. So we can rewrite formula
\eqref{F:J(g,y)} as
\begin{equation}\label{F:J}
\varphi^*(dx)=J(y)d\mu dy
\end{equation}
where $J\in C^\infty(Y)$ with $J\geq 0$.
\end{proposition}

\begin{proof}
We denote the natural action of $h\in G$ on $G/K$ also by $l_h$,
then one can easily verify that
$\sigma_h\circ\varphi=\varphi\circ(l_h\times id)$. By the
$G$-invariance of $dx$ and $d\mu$, we have
\begin{align*}
&J([g],y)d\mu dy\\
=&\varphi^*(dx)\\
=&\varphi^*\circ\sigma_h^*(dx)\\
=&(l_h\times id)^*\circ\varphi^*(dx)\\
=&(l_h\times id)^*(J([g],y)d\mu dy)\\
=&J(h[g],y)(l_h^*(d\mu)\times id^*(dy))\\
=&J([hg],y)d\mu dy.
\end{align*}
So $J([g],y)=J([hg],y)$ for all $g,h\in G$, which means $J$ is
independent of the first variable.
\end{proof}

\begin{corollary}\label{C:J}
There exists a quasi-smooth measure $d\nu$ on $Y$ such that
\begin{equation}\label{F:dmudnu}
\varphi^*(dx)=d\mu d\nu.
\end{equation}
The measure $d\nu$ is given by
\begin{equation}\label{F:dnu}
d\nu(y)=J(y)dy.
\end{equation}
\end{corollary}\qed

The factor $J(y)$ can also be given by more general smooth
measures $u(x)dx$ and $v(y)dy$ on $X$ and $Y$. A direct
calculation yields the following

\begin{proposition}\label{P:bian}
Suppose conditions (a), (b), and (c) hold. If the measures vary by
$dx'=u(x)dx$, $dy'=v(y)dy$, and $d\mu'=\lambda d\mu$, where $u, v$
are positive smooth functions on $X, Y$, respectively, $u$ is
$G$-invariant, and $\lambda$ is a positive constant, then $J(y)$
varies by
$$J'(y)=\frac{u(y)}{\lambda v(y)}J(y).$$
\end{proposition}\qed

Now we suppose that there is a Riemannian structure on $X$ such
that $dx$ and $dy$ are the induced Riemannian measures on $X$ and
$Y$, respectively. We suppose the following orthogonality
condition holds

\vskip 0.3cm {\flushleft{\bf(d)}} \quad (\emph{orthogonality
condition}) \quad $T_yY\perp T_yO_y$, \quad $\forall y\in Y'$.

\vskip 0.3cm {\flushleft Then} we can compute the factor $J(y)$ in
a simple way by the following theorem.

\vskip 0.3cm Let $\Ll$ be a linear subspace of the Lie algebra
$\Lg$ of $G$ such that $\Lg=\Lk\oplus\Ll$, where $\Lk$ is the Lie
algebra of $K$. Let $\pi:G\rightarrow G/K$ be the natural
projection, then $(d\pi)_e|_\Ll:\Ll\rightarrow T_{[e]}(G/K)$ is an
isomorphism. We endow a Riemannian structure on $G/K$ such that
the associated Riemannian measure is $d\mu$, then it also induces
an inner product on $T_{[e]}(G/K)$. For $y\in Y$, we define a
linear map $\Psi_y:\Ll \rightarrow T_yO_y$ by
\begin{equation}\label{E:Psi}
\Psi_y(\xi)=\frac{d}{dt}\Big |_{t=0}\sigma_{\exp t\xi}(y) , \quad
\forall \xi\in \Ll .
\end{equation}
If $y\in Y'$, then $\dim\Ll=\dim T_y O_y$. We choose an inner
product on $\Ll$, and endow the inner product on $T_y O_y$ induced
from the Riamannian structure on $X$. Then the ``determinants"
$|\det\Psi_y|$ and $|\det((d\pi)_e|_\Ll)|$ make sense.

\begin{theorem}\label{T:det}
Under the above assumptions, we have
\begin{equation}\label{F:J=det}
J(y)=C|\det\Psi_y|,
\end{equation}
for $y\in Y'$, where $C=|\det((d\pi)_e|_\Ll)|^{-1}$ is a constant.
\end{theorem}

\begin{proof}
By the transversality condition (b), the tangent map
$$(d\varphi)_{([e],y)}:T_{([e],y)}(G/K\times Y)\rightarrow T_yX$$ of
$\varphi$ at the point $([e], y)$ ($y\in Y'$) can be regarded as
$$
(d\varphi)_{([e],y)}:T_{[e]}(G/K) \oplus T_yY \rightarrow T_yO_y
\oplus T_yY .
$$
Denote
$\widetilde{\Psi}_y=(d\varphi)_{([e],y)}|_{T_{[e]}(G/K)}:T_{[e]}(G/K)\rightarrow
T_yO_y$, then it is obvious that
$\Psi_y=\widetilde{\Psi}_y\circ(d\pi)_e|_\Ll$, and one can easily
show that in the matrix form,
$$(d\varphi)_{([e],y)}=\left(\begin{array}{cc}\widetilde{\Psi}_y&0\\0&id
\end{array}\right).$$

\vskip 0.3cm Since $d\mu$ is the associated Riemannian measure on
$G/K$, the product measure $d\mu dy$ is the the associated
Riemannian measure on the product Riemannian manifold $G/K\times
Y'$. By Proposition \ref{P:pullback} and the orthogonality
condition (d),
\begin{align*}
J(y)=&\left|\det(d\varphi)_{([e],y)}\right|\\
=&\left|\det\left(\begin{array}{cc}\widetilde{\Psi}_y&0\\0&id
\end{array}\right)\right|\\
=&|\det\widetilde{\Psi}_y|\\
=&|\det(\Psi_y\circ((d\pi)_e|_\Ll)^{-1})|\\
=&C|\det\Psi_y|,
\end{align*}
where $C=|\det((d\pi)_e|_\Ll)|^{-1}$. This proves the theorem.
\end{proof}

\begin{remark}
Although formula \eqref{F:J=det} only hold on $Y'$, since $Y'$ is
dense in $Y$ and $J\in C^\infty(Y)$, we can get $J(y)$ for all
$y\in Y$ by smooth continuation.
\end{remark}

\vskip 0.5cm
\section{Integrations over $G$-spaces}

\vskip 0.5cm Occasionally we will be interested in some kinds of
integration formulae. In this section we give some preliminaries
on integrations. The reader who has more interest in the
eigenvalue distributions of the generalized random matrix
ensembles may skip this section and go to \S 4 directly.

\vskip 0.3cm The following proposition generalizes the change of
variables formula for multiple integration.

\begin{proposition}\label{P:jifenbianhuan}
Let $\varphi: M\rightarrow N$ be a smooth map between two
$n$-dimensional smooth manifolds $M$ and $N$, $dy$ a smooth
measure on $N$. If $\varphi$ is a local diffeomorphism and is a
$d$-sheeted covering map, then for any $f\in C^\infty(N)$ with
$f\geq0$ or with $f\in L^1(N,dy)$, we have
\begin{equation}\label{F:jifenbianhuan}
\int_N f(y) dy=\frac{1}{d}\int_M f(\varphi(x)) \varphi^*(dy).
\end{equation}
\end{proposition}

\begin{proof}
It is a standard argument using partition of unity, the details is
omitted here.
\end{proof}

\begin{remark}
Formula \eqref{F:jifenbianhuan} seems like a formula which relates
degree of a map and integration of volume forms on manifold. When
$M, N$ are compact and oriented, then under the conditions of
Proposition \ref{P:jifenbianhuan}, up to a ``$\pm$" sign, formula
\eqref{F:jifenbianhuan} says nothing but of this. But, in general,
the integration of differential forms is not suitable for us. What
we will need is a change of variables formula which should ignore
the negative sign.
\end{remark}

As in the previous section, Let $X$ be a $G$-space, where $X$ is
an $n$-dimensional smooth manifold, $G$ is a Lie group. Then we
have the reduced map $\varphi: G/K\times Y\rightarrow X$. Suppose
$dx, dy$, and $d\mu$ are smooth measures on $X,Y$, and $G/K$,
respectively, with $dx$ and $d\mu$ to be $G$-invariant. Our goal
is to convert the integration over $X$ to the integration over
$Y$. Suppose the conditions (a), (b), and (c) hold. We hope the
map $\varphi: G/K\times Y'\rightarrow X'$ satisfies the conditions
as given in Proposition \ref{P:jifenbianhuan}.

\vskip 0.3cm
\begin{proposition}\label{P:localdiffeo}
Suppose conditions (a), (b) and (c) hold. Then $\varphi: G/K\times
Y'\rightarrow X'$ is a local diffeomorphism.
\end{proposition}

\vskip 0.3cm
\begin{proof} \hskip 0.2cm
Let $e$ be the unit element in $G$. For $([e],y)\in G/K\times Y'$,
$d\varphi_{([e],y)}(0,v)=v, \forall v\in T_y Y'$, so $T_y
Y'\subset Im(d\varphi_{([e],y)})$. Furthermore,
$\varphi|_{G/K\times\{y\}}:G/K\times\{y\}\rightarrow O_y\cong
G/G_y$ is a local diffeomorphism, so $T_y O_y\subset
Im(d\varphi_{([e],y)})$. Thus $d\varphi_{([e],y)}$ is surjective.
But $\mathrm{dim}(G/K\times Y')=\mathrm{dim}X'$, so
$d\varphi_{([e],y)}$ is in fact an isomorphism. For general
$([g],y)\in G/K\times Y'$, notice that $\varphi\circ
l_g=\sigma_g\circ\varphi$, where $l_g([h],y)=([gh],y)$, so
$d\varphi_{([g],y)}\circ
(dl_g)_{([e],y)}=(d\sigma_g)_{([e],y)}\circ d\varphi_{([e],y)}$,
and $d\varphi_{([e],y)}$ is isomorphic implies
$d\varphi_{([g],y)}$ is isomorphic. Thus $\varphi$ is everywhere
regular, and hence is a local diffeomorphism.
\end{proof}

To make Proposition \ref{P:jifenbianhuan} available, we endow the
following covering condition.

\vskip 0.3cm {\flushleft{\bf(e)}} \quad (\emph{covering
condition}) \quad The map $\varphi: G/K\times Y'\rightarrow X'$ is
a $d$-sheeted covering map, with $d<+\infty$.

\vskip 0.3cm
\begin{theorem}\label{T:Weylbianhuan}
Suppose conditions (a), (b), (c), and (e) hold. Then we have
\begin{equation}\label{F:Weylbianhuan}
\int_X f(x)
dx=\frac{1}{d}\int_Y\left(\int_{G/K}f(\sigma_g(y))d\mu([g])\right)
J(y)dy
\end{equation}
for all $f\in C^\infty(X)$ with $f\geq0$ or with $f\in L^1(X,dx)$,
where $J\in C^\infty(Y)$ is determined by Formula (\ref{F:J}).
\end{theorem}

\vskip 0.3cm
\begin{proof}
By Proposition \ref{P:localdiffeo}, $\varphi: G/K\times
Y'\rightarrow X'$ is a local diffeomorphism. By the covering
condition (e), $\varphi$ is a $d$-sheeted covering map. So by
Proposition \ref{P:jifenbianhuan}, for $f\in C^\infty(X)$ with
$f\geq0$ or $f\in L^1(X,dx)$, we have
\begin{align*}
\int_X f(x)dx
=&\int_{X'} f(x)dx\\
=&\frac{1}{d}\int_{G/K\times Y'}f(\varphi([g],y))\varphi^*(dx)\\
=&\frac{1}{d}\int_{G/K\times Y'}f(\sigma_g(y))J(y)d\mu([g])dy\\
=&\frac{1}{d}\int_{Y'}\left(\int_{G/K}f(\sigma_g(y))d\mu([g])\right)J(y)dy\\
=&\frac{1}{d}\int_Y\left(\int_{G/K}f(\sigma_g(y))d\mu([g])\right)J(y)dy
\end{align*}\end{proof}

\vskip 0.3cm
\begin{corollary}\label{C:leihanshubianhuan}
Under the same conditions as in the above Theorem, if furthermore
$f(\sigma_g(x))=f(x), \forall g\in G, x\in X$, then
\begin{equation}\label{F:leihanshubianhuan}
\int_X f(x) dx=\frac{\mu(G/K)}{d}\int_Y f(y)J(y)dy.
\end{equation}
\end{corollary}\qed

\vskip 0.3cm To make the above conclusion more available, we give
some criterions on when the map $\varphi: G/K\times Y'\rightarrow
X'$ is a covering map.

\vskip 0.3cm
\begin{proposition}\label{P:covering} Let $M,N$ be smooth
$n$-dimensional manifolds. Then an everywhere regular smooth map
$\varphi : M\rightarrow N$ is a $d$-sheeted covering map if and
only if for each $y\in N$, $\varphi^{-1}(y)$ has $d$ points.
\end{proposition}

\vskip 0.3cm
\begin{proof} \hskip 0.2cm
The ``$\Rightarrow$" part is obvious. We prove the ``$\Leftarrow$"
part.

\vskip 0.3cm For $y\in N$, let
$\varphi^{-1}(y)=\{x_1,\cdots,x_d\}$. Since $\varphi$ is
everywhere regular, there exists open neighborhood $U_i$ of $x_i$,
$i=1,\cdots,d$, such that $U_i\cap U_j=\emptyset$ for $i\neq j$,
and $\varphi_i=\varphi|_{U_i}:U_i\rightarrow\varphi(U_i)$ is a
diffeomorphism. Let $V=\bigcap_{i=1}^d\varphi(U_i)$, and let
$V_i=\varphi_i^{-1}(V)$, then $\varphi|_{V_i}$ is also a
diffeomorphism onto $V$. We conclude that
$\varphi^{-1}(V)=\bigcup_{i=1}^d V_i$. In fact, $\forall
z\in\varphi^{-1}(V)$, let $z_i=\varphi_i^{-1}(\varphi(z))$, then
 $z_i\in\varphi^{-1}(\varphi(z))$ and $z_i\neq z_j$ for $i\neq j$.
But  $z\in\varphi^{-1}(\varphi(z))$ and
$|\varphi^{-1}(\varphi(z))|=d$, this force $z=z_{i_0}$ for some
$i_0$. Hence $z\in\bigcup_{i=1}^d V_i$. Therefore
$\varphi^{-1}(V)=\bigcup_{i=1}^d V_i$. The Lemma is proved.
\end{proof}

\vskip 0.3cm
\begin{corollary}\label{C:covering}
Suppose conditions (a), (b) and (c) hold. If furthermore $\exists
d\in \mathbb{N}$ such that $\forall y\in Y'$,

\vskip 0.3cm (1) \quad the isotropic subgroup \hskip 0.1cm
$G_y=K$, \hskip 0.6cm (2) \quad $|O_y\cap Y'|=d$,

\vskip 0.3cm \noindent then  $\varphi: G/K\times Y'\rightarrow X'$
is a $d$-sheeted covering map.
\end{corollary}

\vskip 0.3cm
\begin{proof}
By Proposition \ref{P:localdiffeo}, $\varphi$ is a local
diffeomorphism. So by the above Proposition, we need only to show
that for each $x\in X'$, $\varphi^{-1}(x)$ has $d$ points.

\vskip 0.3cm For $x\in Y'$, suppose $O_x\cap
Y'=\{y_1,\cdots,y_d\}$. Then there exists $g_i\in G$ such that
$\sigma_{g_i}(y_i)=x$ for each $i\in\{1,\cdots,d\}$. Then
$([g_i],y_i)\in\varphi^{-1}(x)$. On the other hand, if
$([g],y)\in\varphi^{-1}(x)$, then $y=y_{i_0}$ for some
$i_0\in\{1,\cdots,d\}$. Now we have
$\sigma_{gg_{i_0}^{-1}}(x)=\sigma_{g}(y_{i_0})=x$, that is
$gg_{i_0}^{-1}\in G_x=K$, so $[g]=[g_{i_0}]$ and
$([g],y)=([g_{i_0}],y_{i_0})$. Thus
$\varphi^{-1}(x)=\{([g_1],y_1),\cdots,([g_d],y_d)\}$.

\vskip 0.3cm In general for $x\in X'$, suppose $\sigma_h(x)\in Y'$
for some $h\in G$, then the relation
$\varphi^{-1}(\sigma_{h}(x))=l_h(\varphi^{-1}(x))$ reduces the
general case to the above one.
\end{proof}

\vskip 0.3cm Both Proposition \ref{P:covering} and Corollary
\ref{C:covering} will be used in the following sections when we
consider concrete examples.

\vskip 0.3cm
\begin{remark}
The converse of Corollary \ref{C:covering} is not true. That is,
the isotropic subgroups $G_y$ associated with $y\in Y'$ may vary
``suddenly", even if $Y'$ is connected. For example, The group
$SO(n)$ acts on $\mathbb{R}P^n$ smoothly if we regard
$\mathbb{R}P^n$ as the quotient space by gluing the opposite
points on the boundary of the closed unit ball $B^n$. Let $X_\z$
be the image of $\{0\}$, $Y$ be the image of the segment
$\{(x,0,\cdots,0):|x|\leq 1\}$, then the conditions (a), (b), (c),
and (e) hold. The isotropic subgroup associated with the image of
a point in $Y'$ which is an interior point of $B^n$ is
$\diag(1,SO(n-1))$, but for the image of the point
$(1,0,\cdots,0)$, its isotropic subgroup is
$\diag(\pm1,O^\pm(n-1))$ (here $O^\pm(n-1)=\{g\in O(n-1):\det
g=\pm1\}$ ). Other examples with the similar phenomena will
appeared in \cite{AWY} when we consider the group ensemble
associated with complex semisimple Lie groups. When the phenomena
of sudden variation of the isotropic subgroups happens, whether we
can in general make them to be of the same by enlarging the set
$X_\z$ is an open problem.
\end{remark}

\vskip 0.5cm
\section{Generalized random matrix ensembles}

\vskip 0.5cm Now we are prepared to establish the generalized
random matrix ensembles.

\vskip 0.3cm Let $G$ be a Lie group which acts on an
$n$-dimensional smooth manifold $X$ by $\sigma: G\times
X\rightarrow X$. For the convenience, we suppose $X$ is a
Riemannian manifold. Suppose the induced Riemannian measure $dx$
is $G$-invariant (note that we do not require the Riemannian
structure on $X$ to be $G$-invarinant). Let $Y$ be a closed
submanifold of $X$ which is endowed the induced Riemannian measure
$dy$, and let $K=\{g\in G:\sigma_g(y)=y, \forall y\in Y\}$. As in
\S 2, we form the map $\varphi: G/K\times Y\rightarrow X$ by
$\varphi([g],y)=\sigma_g(y)$. Let $X_\z \subset X$, $Y_\z \subset
Y$ be closed zero measure subsets of $X$ and $Y$, respectively.
Denote $X' = X \setminus X_\z$,  $Y' = Y \setminus Y_\z$. We
suppose the conditions (a), (b), (c), and (d) of \S 2 hold. For
the reader's convenience, we list them below.

\vskip 0.3cm {\flushleft{\bf(a)}} \quad (\emph{invariance
condition}) \quad $X' = \D\bigcup_{y\in Y'} O_y$.

\vskip 0.3cm {\flushleft{\bf(b)}} \quad (\emph{transversality
condition}) \quad $T_y X=T_y O_y\oplus T_y Y$, \quad $\forall y\in
Y'$.

\vskip 0.3cm {\flushleft{\bf(c)}} \quad (\emph{dimension
condition}) \quad $\mathrm{dim}G_y=\mathrm{dim}K, \quad \forall
y\in Y'$.

\vskip 0.3cm {\flushleft{\bf(d)}} \quad (\emph{orthogonality
condition}) \quad $T_yY\perp T_yO_y$, \quad $\forall y\in Y'$.

\vskip 0.3cm {\flushleft Suppose} $d\mu$ is a $G$-invariant smooth
measure on $G/K$, and suppose $p(x)$ is a $G$-invariant smooth
function on $X$. Then by Corollary \ref{C:J}, there is a
quasi-smooth measure $d\nu$ on $Y$ such that
\begin{equation}\label{E:distribution}
\varphi^*(p(x)dx)=d\mu d\nu.
\end{equation}

\begin{definition}\label{D:generalized}
Let the conditions and notations be as above. Then the system
$(G,\sigma,X,p(x)dx,Y,dy)$ is called a \emph{generalized random
matrix ensemble}. The manifolds $X$ and $Y$ are called the
\emph{integration manifold} and the \emph{eigenvalue manifold},
respectively. The measure $d\nu$ on $Y$ determined by
\eqref{E:distribution} is called the \emph{generalized eigenvalue
distribution}.
\end{definition}

Recall that in \S 2 we have defined the map $\Psi_y : \Ll
\rightarrow T_yO_y$ by
$$
\Psi_y (\xi ) = \frac{d}{dt}\Big |_{t=0}\hskip 0.1cm \sigma_{\exp
t\xi}(y) , \quad \forall \xi\in \Ll ,
$$
where $\Ll$ is a linear subspace of $\Lg$ such that
$\Lg=\Lk\oplus\Ll$. Thanks to the preliminaries in \S 2, we can
compute the generalized eigenvalue distribution directly according
to the following theorem.

\begin{theorem}\label{T:distribution}
Let $(G,\sigma,X,p(x)dx,Y,dy)$ be a generalized random matrix
ensemble. Then the generalized eigenvalue distribution $d\nu$ is
given by
\begin{equation}\label{E:distribution=}
d\nu(y)=\mathcal{P}(y)dy=p(y)J(y)dy,
\end{equation}
where
\begin{equation}\label{E:factor=det}
J(y)= C |\det\Psi_y|,
\end{equation}
here $C=|\det((d\pi)_e|_\Ll)|^{-1}$.
\end{theorem}

\begin{proof}
This follows directly from Proposition \ref{P:J}, Corollary
\ref{C:J}, Proposition \ref{P:bian} and Theorem \ref{T:det}.
\end{proof}

The function $\mathcal{P}(y)=p(y)J(y)$ determined by formula
\eqref{E:distribution=} is called the \emph{generalized joint
density function}.

\vskip 0.3cm One of the most fundamental problems in the random
matrix theory is to compute the eigenvalue distribution $d\nu$. In
our generalized scheme, it is given by formulae
\eqref{E:distribution=} and \eqref{E:factor=det}. Note that the
power of \eqref{E:factor=det} is reflected by the fact that it
provides a direct and unified method to compute the eigenvalue
distributions of various kinds of random matrix ensembles. In the
sequel paper \cite{AWY}, we will see that all the classical
ensembles are included in the generalized scheme, and the
corresponding eigenvalue distributions can be derived from
\eqref{E:distribution=} and \eqref{E:factor=det}. We will also
present various kinds of generalized ensemble which were not
considered before, and compute their eigenvalue distributions
explicitly.

\vskip 0.3cm Now we consider the integration formula associated
with the generalized random matrix ensemble. As in \S 3, we assume
the following covering condition holds.

\vskip 0.3cm {\flushleft{\bf(e)}} \quad (\emph{covering
condition}) \quad The map $\varphi: G/K\times Y'\rightarrow X'$ is
a $d$-sheeted covering map, with $d<+\infty$.

\begin{theorem}\label{T:random-integration}
Let $(G,\sigma,X,p(x)dx,Y,dy)$ be a generalized random matrix
ensemble. Suppose the covering condition (e) holds. Then we have
the following integration formula
\begin{equation}\label{E:integration}
\int_X f(x)p(x)
dx=\frac{1}{d}\int_Y\left(\int_{G/K}f(\sigma_g(y))d\mu([g])\right)
d\nu(y)
\end{equation}
for all $f\in C^\infty(X)$ with $f\geq0$ or with $f\in
L^1(X,p(x)dx)$. If moreover $f(\sigma_g(x))=f(x), \forall g\in G,
x\in X$, then
\begin{equation}\label{E:classintegration}
\int_X f(x)p(x) dx=\frac{\mu(G/K)}{d}\int_Y f(y)d\nu(y).
\end{equation}
\end{theorem}

\begin{proof}
It is obvious by Theorem \ref{T:Weylbianhuan} and Corollary
\ref{C:leihanshubianhuan}.
\end{proof}

In formula \eqref{E:classintegration}, if the measure $p(x)dx$ is
a probability measure, and we let $f=1$, we get
$\frac{\mu(G/K)}{d}\int_Y d\nu(y)=1$. So if $G/K$ is compact, we
can normalized the measure $d\mu$ such that $\mu(G/K)=d$, then the
generalized eigenvalue distribution $d\nu$ is a probability
measure.

\begin{remark}
The condition $f\in C^\infty(X)$ in Theorem
\ref{T:random-integration} is superfluous. In fact, it is
sufficient to assume $f$ is measurable. The same is true for
Proposition \ref{P:jifenbianhuan} and Theorem
\ref{T:Weylbianhuan}.
\end{remark}

\vskip 0.5cm
\section{A classification scheme of generalized ensembles}

\vskip 0.5cm In this section we give a classification scheme of
the generalized random matrix ensembles, that is, \vskip 0.2cm
{\flushleft(1)} Linear ensemble,\\
(2) Nonlinear noncompact ensemble,\\
(3) Compact ensemble,\\
(4) Group ensemble,\\
(5) Algebra ensembles,\\
(6) Pseudo-group ensemble,\\
(7) Pseudo-algebra ensemble.

\vskip 0.3cm First we define the linear ensemble and the nonlinear
noncompact ensemble. Let $G$ be a real reductive Lie group with
Lie algebra $\Lg$ in the sense of Knapp \cite{Kn}, Section 7.2.
Then $G$ admits a global Cartan involution $\Theta$, which induces
a Cartan involution $\theta$ of $\Lg$. Let the corresponding
Cartan decomposition of $\Lg$ is $\Lg=\Lk\oplus\Lp$. Let $K=\{g\in
G:\Theta(g)=g\}$, $P=\exp(\Lp)$, then $K$ is a maximal compact
subgroup of $G$ with Lie algebra $\Lk$, $P$ is a closed
submanifold of $G$ satisfies $T_eP=\Lp$. The spaces $\Lp$ and $P$
are invariant under the adjoint action $\A=\Ad|_K$ and the
conjugate action $\sigma$ of $K$, respectively. Let $\La$ be a
maximal abelian subspace of $\Lp$, and let $A$ be the connected
subgroup of $G$ with Lie algebra $\La$. Let $M=\{k\in
K:\A_k(\eta)=\eta, \forall \eta\in\La\}=\{k\in K:\sigma_k(a)=a,
\forall a\in A\}$. It can be shown that there are Riemannian
structures on $\Lp$ and $P$ inducing $K$-invariant Riemannian
measures $dX$ on $\Lp$ and $dx$ on $P$. They also induce
Riemannian measures $dY$ on $\La$ and $da$ on $A$. There is also a
$K$-invariant smooth measure $d\mu$ on $K/M$. Let $p_1(\xi)$ and
$p_2(x)$ be $K$-invariant positive smooth functions on $\Lp$ and
$P$, then it can be proved that the systems
$(K,\A,\Lp,p_1(\xi)dX(\xi),\La,dY)$ and
$(K,\sigma,P,p_2(x)dx,A,da)$ are generalized random matrix
ensembles, which we called \emph{linear ensemble} and
\emph{nonlinear noncompact ensemble}, respectively. It can be
shown that the Gaussian ensemble and the chiral ensemble are
particular examples of linear ensemble, and the new transfer
matrix ensembles are particular examples of nonlinear noncompact
ensemble.

\vskip 0.3cm Next we define the compact ensemble. Let $G$ be a
connected compact Lie group $G$ with Lie algebra $\Lg$. Suppose
$\Theta$ is a global involutive of $G$ with the induced involution
$\theta=d\Theta$ of $\Lg$. Let $K=\{g\in G:\Theta(g)=g\}$, and let
$\Lp$ be the eigenspace of $\theta$ corresponding the eigenvalue
$-1$. Let $P=\exp(\Lp)$, then $P$ is invariant under the conjugate
action $\sigma$ of $K$. It was proved in \cite{AW} that $P$ is a
closed submanifold of $G$ satisfies $T_eP=\Lp$, which is just the
identity component of the set $\{g\in G:\Theta(g)=g^{-1}\}$. Let
$\La$ be a maximal abelian subspace of $\Lp$, and let $A$ be the
torus with Lie algebra $\La$. There is a Riemannian structure on
$P$ induces a $K$-invariant Riemannian measure $dx$ on $P$ and a
Riemannian measure $da$ on $A$. Let $M=\{k\in K:\sigma_k(a)=a,
\forall a\in A\}$, then there is a $K$-invariant smooth measure
$d\mu$ on $K/M$. Let $p(x)$ be a $K$-invariant positive smooth
function on $P$, then it can be proved that the system
$(K,\sigma,P,p(x)dx,A,da)$ is a generalized random matrix
ensemble, which we call it \emph{compact ensemble}. It can be
shown that the circular ensemble and the Jacobi ensembles are
particular examples of compact ensemble.

\vskip 0.3cm Let $G$ be an unimodular Lie group $G$ with Lie
algebra $\Lg$. Then there are Riemannian structures on $G$ and
$\Lg$ inducing a $\sigma$-invariant Riemannian measure $dg$ on $G$
and an $\Ad$-invariant Riemannian measure $dX$ on $\Lg$, where
$\sigma$ denotes the conjugate action of $G$ on itself. Let
$p_1(g)$ and $p_2(\xi)$ be two function on $G$ and $\Lg$,
respectively, which are invariant under the corresponding actions
of $G$. If there exists a closed submanifold $Y$ of $G$ such that
$(G,\sigma,G,p(g)dg,Y,dy)$ is a generalized random matrix
ensemble, where $dy$ is the induced Riemannian measure on $Y$,
then we call it a \emph{group ensemble}. And if there exists a
closed submanifold $\Ly$ of $\Lg$ such that
$(G,\Ad,\Lg,p(\xi)dX(\xi),\Ly,dY)$ is a generalized random matrix
ensemble, where $dY$ is the induced Riemannian measure on $\Ly$,
then we call it an \emph{algebra ensemble}. Among all the
unimodular Lie groups, the connected compact Lie group and the
connected complex semisimple Lie group are of particular interest.
For a connected compact Lie group $G$, we can let the submanifold
$Y$ of $G$ be a maximal torus $T$ of $G$, and let the submanifold
$\Ly$ of $\Lg$ be the Lie algebra of $T$. For a connected complex
semisimple Lie group $G$, we can let the submanifold $\Ly$ of
$\Lg$ be a Cartan subalgebra of $\Lg$, and let the submanifold $Y$
of $G$ be the corresponding Cartan subgroup of $G$. For these
cases, it can be proved that the systems
$(G,\sigma,G,p(g)dg,Y,dy)$ and $(G,\Ad,\Lg,p(\xi)dX(\xi),\Ly,dY)$
are generalized random matrix ensembles.

\vskip 0.3cm Now we define the pseudo-group ensemble and the
pseudo-algebra ensembles, which are related to real reductive
groups. Let $G$ be a real reductive group with lie algebra $\Lg$.
Let $\theta$ be a Cartan involution of $\Lg$, and let
$\Lh_1,\cdots,\Lh_m$ be a maximal set of mutually nonconjugate
$\theta$ stable Cartan subalgebras of $\Lg$ with the corresponding
Cartan subgroups $H_1,\cdots,H_m$ of $G$. Denote the sets of all
regular elements in $G$ and $\Lg$ by $G_r$ and $\Lg_r$. Let
$H'_j=H_j\cap G_r$, $\Lh'_j=\Lh_j\cap\Lg_r$. Then it is known that
$G_r=\bigsqcup_{j=1}^m\bigcup_{g\in G}\sigma_g(H'_j)$ (see
\cite{Kn}, Theorem 7.108), $\Lg_r=\bigsqcup_{j=1}^m\bigcup_{g\in
G}\Ad_g(\Lh'_j)$ (see \cite{Wa}, Proposition 1.3.4.1), here the
symbol ``$\bigsqcup$" means disjoint union. Each $\bigcup_{g\in
G}\sigma_g(H'_j)$ is an open set in $G$, and each $\bigcup_{g\in
G}\Ad_g(\Lh'_j)$ is an open set in $\Lg$. Let
$G_j=\overline{\bigcup_{g\in G}\sigma_g(H'_j)}$,
$\Lg_j=\overline{\bigcup_{g\in G}\Ad_g(\Lh'_j)}$. It can be shown
that some suitable Riemannian structures on $G$ and $\Lg$ induce a
$\sigma$-invariant measure $dg_j$ on $G_j$ and an $\Ad$-invariant
measure $dX_j$ on $\Lg_j$ for each $j$, and they also induce a
Riemannian measure $dh_j$ on $H_j$ and a Riemannian measure $dY_j$
on $\Lh_j$. It is known that $Z(H_j)=\{g\in
G:\sigma_g(h)=h,\forall h\in H_j\}$, $H_j=\{g\in
G:\Ad_g(\xi)=\xi,\forall \xi\in\Lh_j\}$. Let $d\mu'_j, d\mu_j$ be
$G$-invariant measures on $G/Z(H_j)$ and $G/H_j$, respectively. In
general, the spaces $G_j$ and $\Lg_j$ may have singularities. But
this doesn't matter, since they are closures of open submanifolds
in $G$ and $\Lg$, whose boundaries have measure zero. If we ignore
this ambiguity, then it can be proved that
$(G,\sigma,G_j,dg_j,H_j,dh_j)$ and $(G,\Ad,\Lg_j,dX_j,\Lh_j,dY_j)$
are generalized random matrix ensembles, which we called
\emph{pseudo-group ensemble} and \emph{pseudo-algebra ensemble},
respectively.

\vskip 0.3cm Due to the generality of the definition, our
classification could not exhaust all kinds of generalized
ensemble. But it would include all kinds of classical random
matrix ensembles and some new examples of generalized ensembles,
which will be analyzed explicitly in the sequel paper \cite{AWY}.

\end{document}